\documentclass{iaus237}
\usepackage{graphicx}
\usepackage{natbib}  

\title[Sequentially Triggered Star Formation in OB Associations] %% give here short title %%
{Sequentially Triggered Star Formation in OB Associations}

\author[Th.~Preibisch \& H.~Zinnecker]   %% give here short author list %%
{Thomas Preibisch$^1$
  \and Hans Zinnecker$^2$}

\affiliation{$^1$Max-Planck-Institut f\"ur Radioastronomie,
 Auf dem H\"ugel 69, D--53121 Bonn, Germany \break email: preib@mpifr-bonn.mpg.de\\[\affilskip]
$^2$Astrophysikalisches Institut Potsdam, An der Sternwarte 16, 
D--14482
  Potsdam, Germany \break email: hzinnecker@aip.de}

\pubyear{2007}
\volume{237}  %% insert here IAU Symposium No.
\pagerange{119--126}
\date{?? and in revised form ??}
\jname{Triggered Star Formation in a Turbulent ISM}
\editors{B.G. Elmegreen \& J. Palous, eds.}
\begin{document}

\maketitle

\begin{abstract}
We discuss observational evidence for sequential and
triggered star formation in 
OB associations. We first review the star
formation process in the Scorpius-Centaurus OB association,
the nearest OB association to the Sun, where several 
recent extensive studies have 
allowed us to 
reconstruct the star formation history
in a rather detailed way. 
We then compare the observational
results with those obtained for other OB associations and with
recent models of rapid cloud and star formation in the
turbulent interstellar medium.
We conclude that the formation of whole OB subgroups (each 
consisting of several thousand stars) requires large-scale
triggering mechanisms such as shocks from expanding wind and
supernova driven superbubbles surrounding older subgroups.
Other triggering mechanisms, like radiatively driven implosion
of globules, also operate, but 
seem to be  secondary processes, forming only small stellar groups
rather than whole OB subgroups with thousands of stars. 
%% add here a maximum of 10 keywords, to be taken form the file <Keywords.txt>
\end{abstract}

\firstsection % if your document starts with a section,
              % remove some space above using this command.
\section{Star formation in OB associations}

OB associations \citep{Blaauw64}
are loose, co-moving stellar groups containing O- and/or
early B-type stars.  As these associations are
unstable against galactic tidal forces, they
must be young ($\lesssim 30$~Myr)
entities, with most of their low-mass members \citep{Briceno06}
still in their pre-main sequence (PMS) phase.  
OB associations are ideal targets for a detailed 
investigation of the initial mass function (IMF) and the 
star formation history, since they allow us 
to study the {\em outcome of a recently completed star formation
process}.
%%%%%%%%%%%%%%%%%%%%%%
As noted by Blaauw (1964), many OB associations consist of distinct subgroups 
with different ages,
which often seem to progress in a systematic way, 
suggesting a sequential, perhaps triggered formation scenario.

The massive stars in OB associations 
affect their environment by ionizing radiation,
stellar winds, and, finally, supernova explosions.
In their immediate neighborhood, these effects are
mostly destructive, since they tend to disrupt the parental molecular cloud
and thus terminate the star formation process  (e.g. Herbig 1962).
A little further away, however, massive stars
may also stimulate new star formation.
For example,
the ionization front from a massive star can
sweep up surrounding cloud material into a dense shell,
which then fragments and forms new stars 
\citep[the ``collect and collapse'' model; see][]{EL77,Zavagno06}. 
Also, shock waves from 
expanding wind- and/or supernova-driven superbubbles
can trigger cloud collapse and star formation.

A general problem of scenarios of triggered star formation is that
a clear proof of causality is hard to obtain.
One can often see Young Stellar Objects (YSOs) and 
ongoing star formation near shocks caused by massive stars, suggesting 
triggered star formation.
However, such morphological evidence alone
does not constitute unequivocal proof for a triggered star formation scenario.
It is not clear whether the shock really compressed an empty cloud and 
triggered the birth of the YSOs, or whether the YSOs formed before the
shock wave arrived and
the shock appears at the edge
of an embedded star formation site simply because the associated dense cloud
material has slowed it down.
More insight can be gained if one can
determine the ages of the YSOs and compare them to the moment in time
at which the shock arrived. Agreement of these timings
provides much more solid evidence for the triggered
star formation scenario than the spatial alignment alone.

\smallskip

\noindent{\bf Theoretical models for the formation of OB subgroups:}\\
The classical model for the sequential formation of OB subgroups was developed
by Elmegreen \& Lada (1977; see also Lada 1987).
Low-mass stars are assumed to form  spontaneously
throughout the molecular cloud. 
As soon as the first  massive stars form,
their ionizing radiation and winds disperse the cloud in their
immediate surroundings, thereby terminating the local star formation process.
The OB star radiation and winds also drive shocks into other parts of
the cloud. A new generation of massive stars is then formed in the dense shocked layers,
and the whole process is repeated until the wave of propagating star formation
reaches the edge of the cloud. According to this model, one would expect that
(1) the low mass stars should be systematically older than the associated 
OB stars and should show a large age spread, corresponding to the total
lifetime of the cloud, and that (2) the youngest OB subgroups should have the largest
fraction of low-mass stars (as in these regions low-mass star formation 
continued for the longest period of time).
Another model is based on the mechanism of radiation-driven implosion
\citep[e.g.,][]{Kessel03}.
As an OB star drives an ionization shock front into the surrounding cloud,
cores within the cloud are triggered into collapse by the  shock wave.
This model predicts that (1) the low-mass stars should be younger than the
OB stars (which initiate their formation), and (2) one may expect to see
an age gradient in the low-mass population (objects
closer to the OB stars were triggered first and thus should be older than 
those further away; see, e.g.~Chen, these proceedings).

The third model we consider here assumes that a shock wave
driven by stellar winds and/or supernova explosions
runs though a molecular cloud.
Several numerical studies
\citep[e.g.][]{Vanhala98} have found
that the effect of the passing shock wave mainly
depends on the type of the shock and its velocity:
close to a supernova,
the shock wave will destroy ambient clouds, but at larger distances,
when the shock velocities have decreased to
below $\sim 50$~km/s, cloud collapse can be triggered in the
right circumstances.
The distance from the shock source at which the shock properties
are suitable for triggering cloud collapse depend on the details
of the processes creating the
shock wave \citep[see][]{Oey04}, the structure of the surrounding
medium, and the evolutionary state of the pre-impact core, but should
typically range  between
$\sim 20$ pc and $\sim 100$ pc.
This model predicts that (1) low- and high-mass stars in the triggered 
subgroup should have the same age, and (2) the age spread in the new subgroup
is small (since the triggering shock wave crossed the cloud
quite quickly).

These quite distinct predictions of the different models can be compared
to the observed properties (i.e.~the IMF and the star formation history)
of OB associations. The obvious first step of such a study is to identify
the complete or a representative sample of the full stellar population of
the association.
While, at least in the nearby OB associations,
 the population of high- and intermediate mass stars has been
revealed by Hipparcos, the low-mass members (which are usually too 
faint for proper-motion studies) are quite hard to find.
Unlike stellar clusters, which can be
easily recognized on the sky, OB associations are generally not
so conspicuous because they extend over huge areas in the sky (often
several hundred square-degrees for the nearest examples) and most of the faint
stars in the area actually are unrelated fore- or background stars.
Finding the faint low-mass association members among these field stars 
is often like
finding needles in a haystack. However, the availability of powerful
multiple-object spectrographs has now made large
spectroscopic surveys for low-mass  members possible.
The young association members can, e.g., be identified 
by the strength of their 6708~\AA\, Lithium line, 
which is a reliable signature for young stars.
Studies of the {\em complete} stellar population of OB associations
are now feasible and allow us
to investigate in detail the spatial and temporal
relationships between high- and low-mass members.

\section{Triggered star formation in the Scorpius-Centaurus OB association}\label{sec:ScoCen}

At a distance of only $\sim 140$~pc, the Scorpius-Centaurus (ScoCen) 
association is the OB association
nearest to the Sun.  It contains at
least $\sim$150 B stars which concentrate in the three subgroups Upper
Scorpius (USco), Upper Centaurus-Lupus (UCL), and Lower Centaurus-Crux (LCC).
The ages for the B-type stars
in the different
subgroups, derived from the main sequence turnoff in the HR diagram,
were found to be $\sim 5$~Myr for USco, $\sim 17$~Myr for UCL, and
$\sim 16$~Myr for LCC \citep{deGeus89, Mamajek02}.

Upper Scorpius is the best studied part of the ScoCen
complex. 
\Citet{deZeeuw99} identified 120 stars listed in the Hipparcos Catalogue
as genuine members of  high- and intermediate mass
($\sim 20 - 1.5\,M_\odot$). 
After the
first systematic large-scale search for low-mass members 
of USco by \citet{Walter94}, 
we have performed  extensive spectroscopic surveys for further
low-mass members with wide-field
multi-object spectrographs at the Anglo-Australian Observatory.
These observations are
described in detail in \citet{Preibisch98} and 
\citet[][P02 hereafter]{Preibisch02}, 
and ultimately yielded a
sample of 250 low-mass members in the mass range $\sim 0.1\,M_\odot$ to $\sim
2\,M_\odot$.
In combination with the Hipparcos sample of high- and
intermediated mass stars, this large sample allowed
P02 to  study the properties of the full stellar population in USco
on the basis of a statistically robust and well defined sample of members.
The main results of our detailed analysis of this sample
can be summarized as follows:
(1) The stellar mass function in USco is consistent with recent field star
and cluster IMF determinations.
(2) High- as well as low-mass stars have a common mean age of 5~Myr.
(3) The spread seen in the HRD, that may seem to suggest an age spread,
can be fully explained by the effects of the  spread of
individual stellar distances,
unresolved binary companions, and the photometric variablity of the
young stars. The observed HRD provides 
{\em no evidence for an age dispersion}, although small ages spreads of $\sim
1\!-\!2$~Myr cannot be excluded. 
(4) The initial size of the association and the observed internal velocity
dispersion of the members yield a stellar crossing time of $\sim 20$~Myr.

A very important implication of these results is that the observed 
age spread of at most $\leq 1\!-\!2$~Myr is much smaller 
than the stellar crossing time of $\sim$ 20 Myr.
This clearly shows that some
external agent must have coordinated the onset of the star
formation process over the full spatial extent of the association. 
In fact, a very suitable trigger is actually available:
The structure and kinematics of the large
H~I loops surrounding the ScoCen association suggest that
a shock wave from the older UCL group, driven by stellar winds and
supernova explosions, passed through the USco region
just about 5~Myr ago \citep{deGeus92}, which
agrees very well with the ages of the USco members.
\begin{figure}\hspace{1.5cm}\parbox{13cm}{
\parbox{12cm}{ \includegraphics[width=5.0cm]{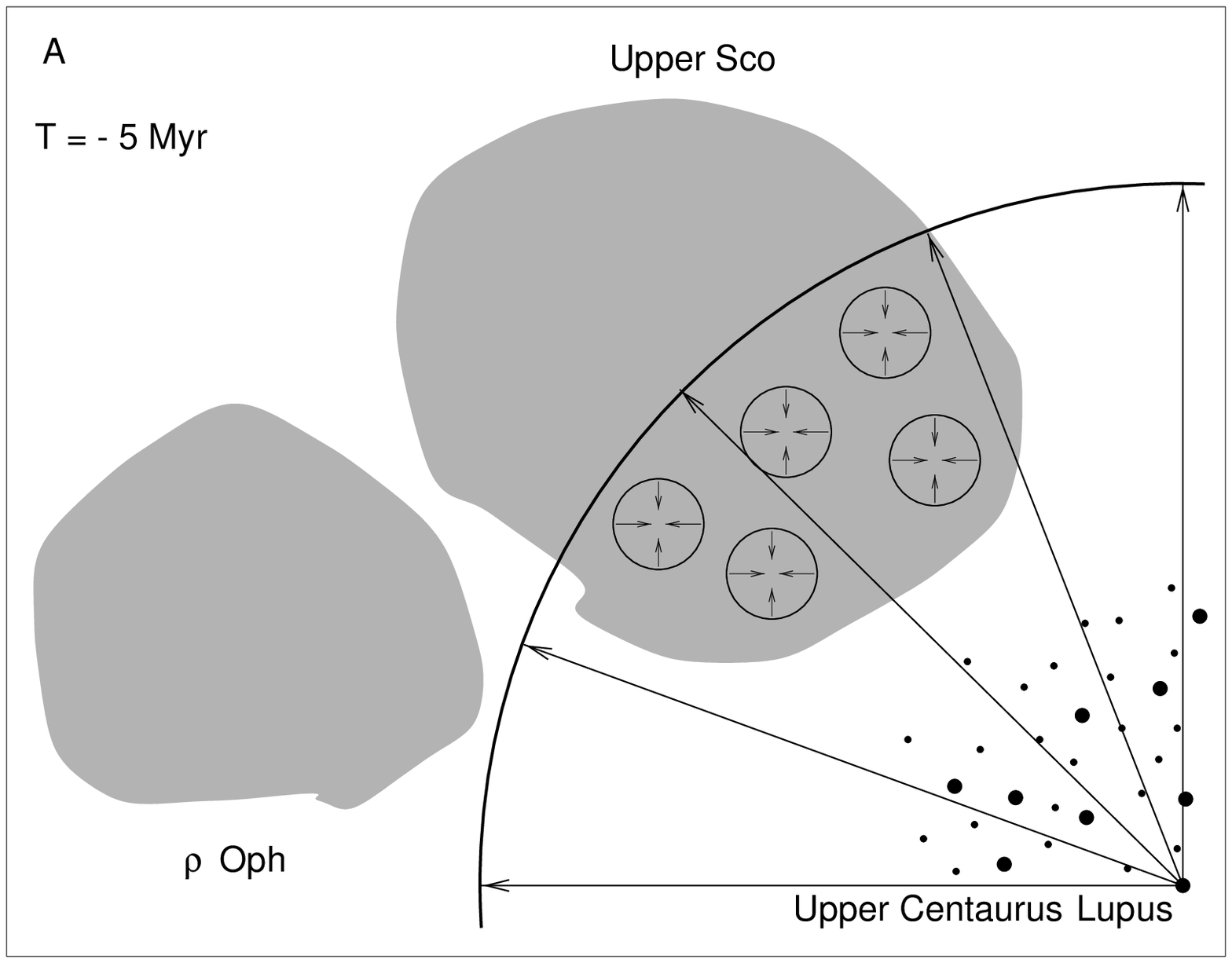}
 \includegraphics[width=5.0cm]{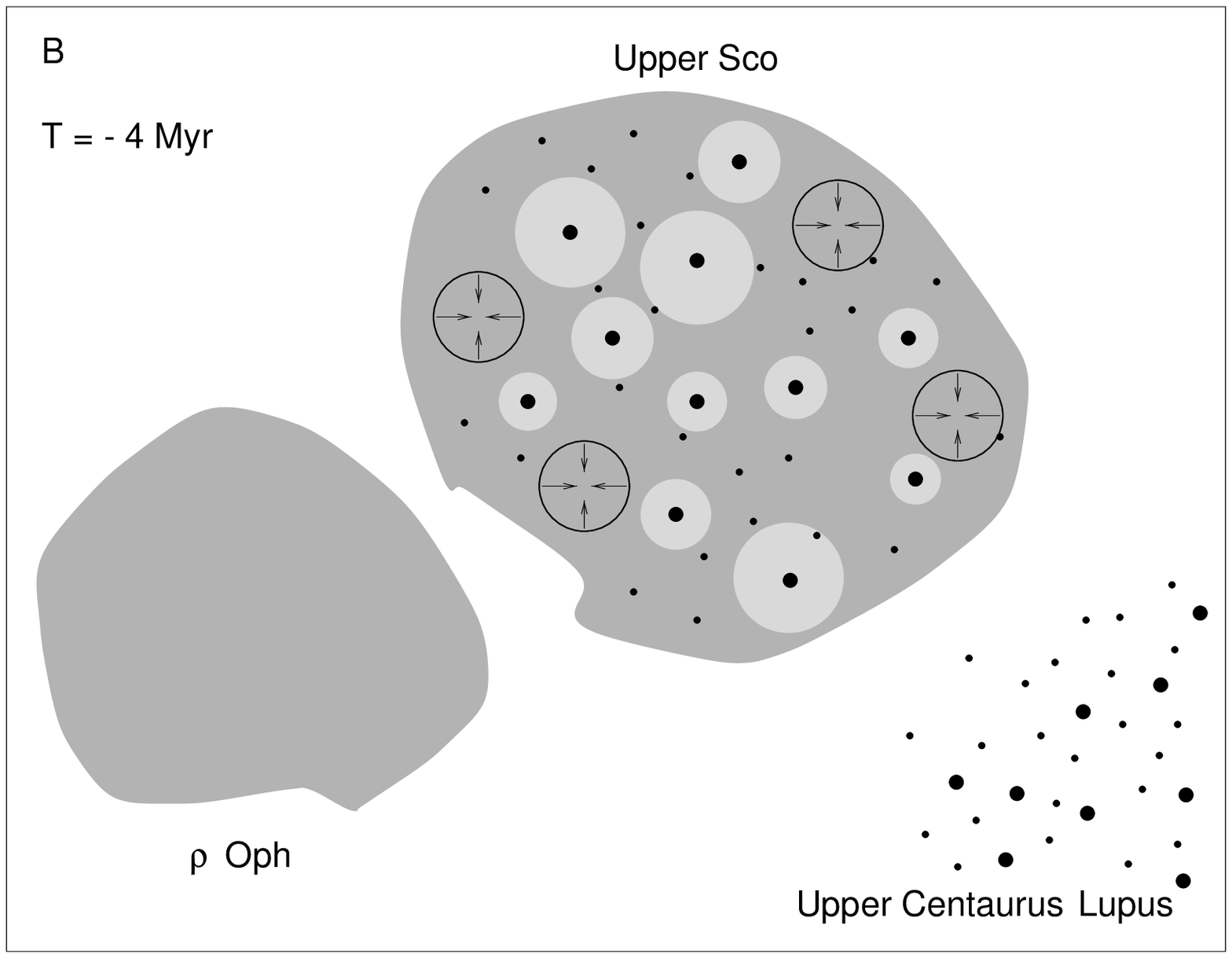}}\newline
\parbox{12cm}{ \includegraphics[width=5.0cm]{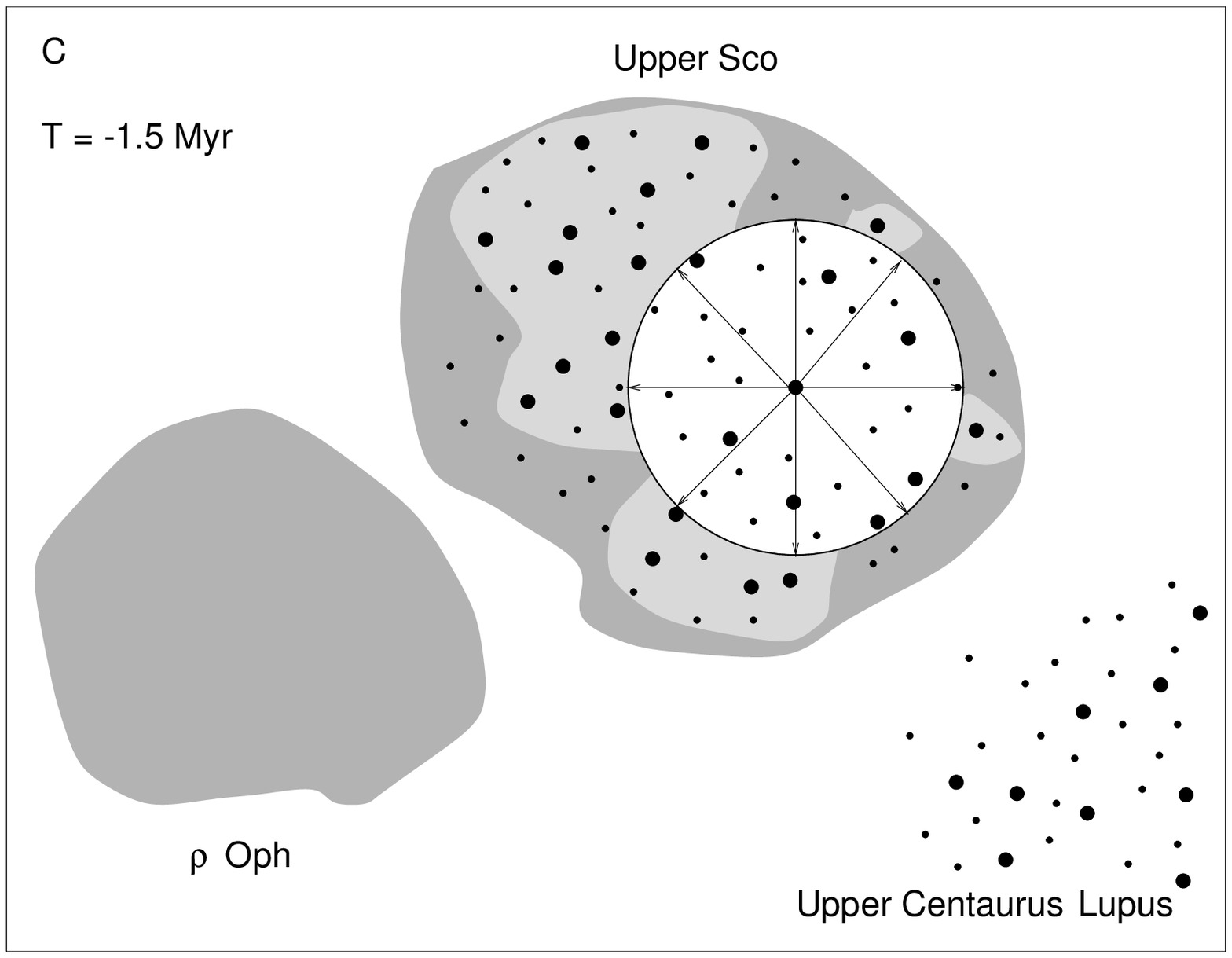}
 \includegraphics[width=5.0cm]{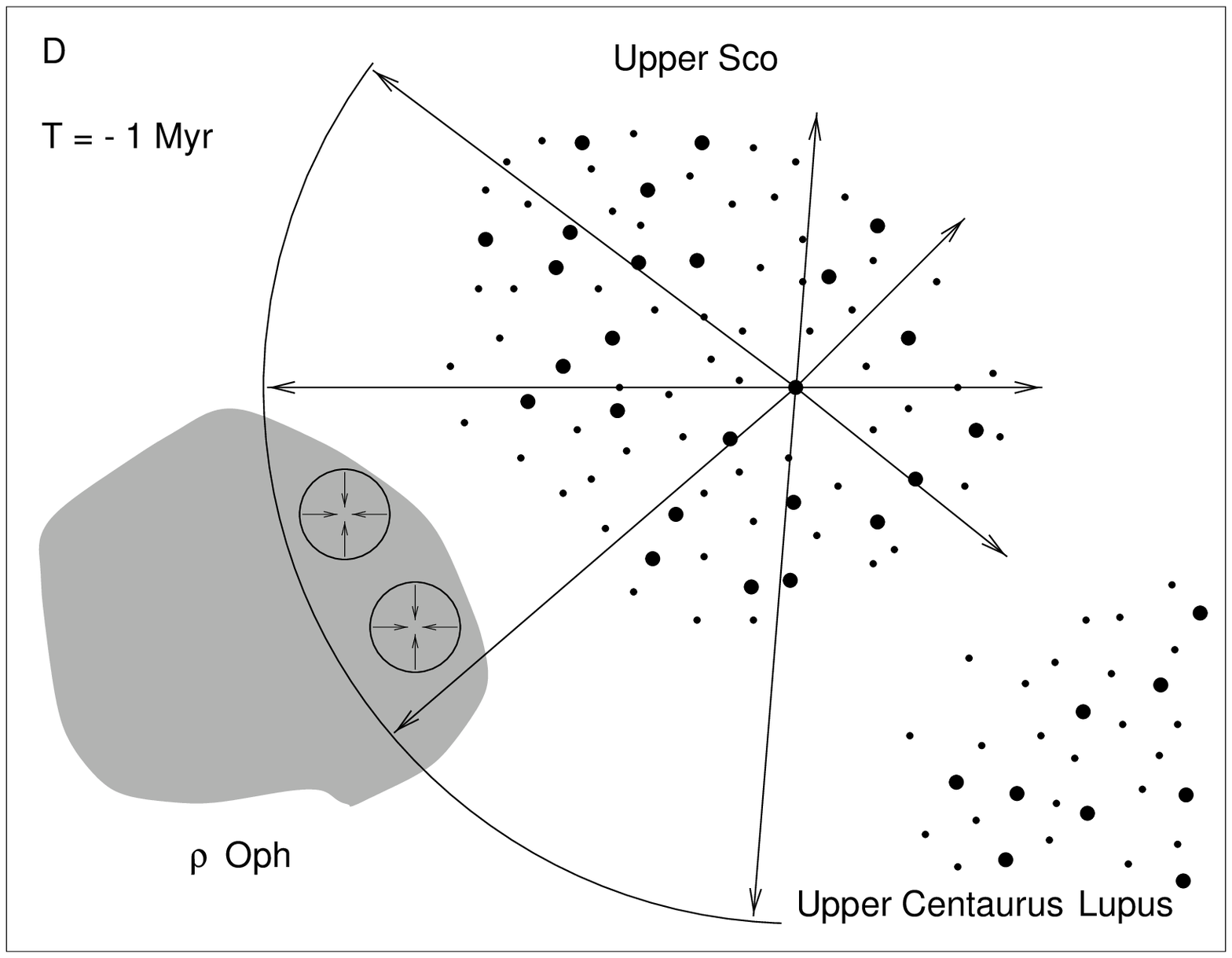}} }
\caption{Schematic view of the star formation history in the
Scorpius-Centaurus association (from Preibisch \& Zinnecker 1999). 
Molecular clouds are shown as dark
regions, high-mass and low-mass stars as large resp.~small dots. For
further details on the sequence of events see text.}
\label{sfm}
\end{figure}
A scenario for the star formation history of USco consistent with
the observational results described above is shown in Fig.~\ref{sfm}.
The shock-wave from UCL initiated the
formation of some 2500 stars in USco, including 10 massive stars upwards of
$10\,M_\odot$.  The new-born massive stars 
immediately started to destroy the cloud from inside by their ionizing
radiation and their strong winds and terminated the 
star formation process; this explains the narrow age distribution.
About 1.5 Myr ago, the most massive star in USco
exploded as a supernova and created another  strong shock
wave, which fully dispersed the USco molecular cloud.
This shock wave must have
crossed the adjacent $\rho$ Oph molecular cloud within the last 1 Myr \citep{deGeus92},
and thus seems to have triggered the
strong star formation activity we witness right now in the
L~1688 cloud \citep[see][]{Motte98}.
 Yet another region in which the shock from USco seems to 
have triggered star formation is the Lupus 1 cloud \citep[see, e.g.,][]{Tachihara01}.

While this scenario provides a good explanation of the star formation history,
a potential problem is its implicit assumption that
the USco and $\rho$ Oph molecular clouds existed for many Myr without
forming stars before the triggering shock waves arrived (otherwise one should
see large age spreads in the stellar populations).
\citet{Elemegreen00}, \citet{Hartmann01}, and other authors provided 
several convincing arguments that the lifetime 
of molecular clouds 
is much shorter than previously thought, 
and the whole process of molecular cloud formation, star
formation, and cloud dispersal (by the feedback of the newly formed stars)
occurs on timescales of a few $(\lesssim 5$)~Myr. 
It is now thought that molecular clouds form by the interaction of
turbulent convergent 
flows in the interstellar medium that accumulate matter in some regions.
Once the H$_2$ column density becomes high enough to provide
effective self-shielding against the ambient UV radiation field,
conversion of atomic H into molecular H$_2$ quickly follows
\citep[e.g.,][]{Glover06}, and star formation may start soon afterwards
\citep[e.g.,][]{Clark05}.

This new paradigm for the formation and lifetime of molecular clouds
may seem to invalidate the idea of a shock wave hitting a
pre-existing molecular cloud and triggering star formation.
Nevertheless, the basic scenario for the sequence of processes in ScoCen
may still be valid. 
As pointed out by \citet{Hartmann01}, wind and
supernova shock waves from massive stars are an important kind of driver for
ISM flows, and are especially well suited to create {\em coherent large-scale 
flows}.
Only large-scale flows are able to form large molecular clouds, in which
whole OB associations can be born.
An updated  scenario for ScoCen could be as follows:
Initially, the winds of the OB stars created an expanding
 superbubble around UCL which interacted 
 with flows in the ambient ISM and 
swept up clouds in some places.
When supernovae started to explode in UCL (note that there were presumably
some 6 supernova explosions in UCL up to the present day), these added energy and 
momentum to the wind-blown superbubble and accelerated its expansion. 
The accelerated shock wave (now with  $v \sim 20 - 30$~km/s) crossed
a swept-up cloud in the USco area, and the increased pressure due to
this shock triggered star formation.

This scenario does not only explain the temporal sequence of events
in a way consistent with the ages of the stars and the kinematic properties
of the observed HI shells.
The following points provide further evidence:
(1) The model \citep[see Fig.~3 in][]{Hartmann01} predicts that stellar groups triggered
in swept-up clouds should be moving away from the trigger source.
A look at the centroid space motions of the ScoCen subgroups \citep{deBruijne99} 
actually shows that USco is moving  away from UCL with a
velocity of $\sim 5 (\pm3)$~km/s.
Furthermore, (2) a study by \citet{Mamajek01} revealed that several
young stellar groups, including the $\eta$~Cha cluster, the TW Hydra association,
and the young stars associated with the CrA cloud,
move away from UCL at velocities of about 10 km/s; tracing their current motions
back in time shows that these groups were located near the edge of UCL 12 Myr ago
(when the supernova exploded). 
Finally, (3)
the model also predicts that molecular clouds are most efficiently
created at the intersection of two expanding bubbles. 
The Lupus I cloud, which is located just between USco and UCL,
may be a good example of this process.
Its elongated shape
is very consistent with the idea that it was swept up by the interaction 
of the expanding superbubble from USco and the (post-SN) bubble created by the 
winds of the remaining early B stars in UCL.

\section{Triggered star formation in other OB associations}

While model and observations agree quite well for ScoCen,
it is important to consider how
general these findings are.
We first note that numerous 
other regions show similar patterns of sequential star formation,
e.g.~the superbubbles in W3/W4a (Oey et al.~2005),  
or the supergiant shell region in IC~2574 (Cannon et al.~2005).
Some OB associations, however, show subgroups with similar ages, 
which cannot have 
formed sequentially. This could nevertheless fit into the proposed scenario,
if we consider the example of Hen 206 in the LMC (Gorjian 2004), where
a supernova driven shock wave from a $\sim 10$~Myr old OB association has
created a huge expanding superbubble, at the southern edge of which  stars are
forming in a giant swept-up molecular cloud. This cloud contains several
discrete peaks, in each of which an entire new OB subgroup seems to be forming.
This could be
an example of the simultaneous formation of several OB subgroups.

A second important aspect is that most nearby OB associations
seem to share some key properties: (1) Their mass function is consistent with
the field IMF, without much evidence for any IMF variations.
(2) High- and low-mass stars generally have the same ages and thus have
formed together, not one first and the other later.
(3) In most regions, age spreads are remarkably small, often much
smaller than the stellar crossing time.
These properties are more consistent with the model of large-scale,
fast triggering by passing shock waves from wind
and supernova driven expanding (super-)bubbles than with the other
triggering mechanisms mentioned above.

Nevertheless, other triggering mechanisms seem to operate simultaneously, 
at least
in some regions. For example, 
the IC 1396 H\,II region in the Cep OB2 association
 contains several globules, which
are strongly irradiated by the central O6 star. Spitzer observations
of the globule VDB 142 (Reach et al. 2004) 
and Chandra X-ray observations of another globule (Getman et al.~2006)
have revealed several very young stellar objects within these globules,
all of which are located close to the illuminated edge of the globules,
providing evidence for triggered star formation. The triggering mechanism
at work here (radiation-driven implosion of globules)
is obviously different from the large-scale triggering by
expanding wind/supernova driven superbubbles. However, 
just a handful of stars are formed in these globules and thus 
this mechanism seems to be a small-scale, secondary
process, which does not seem capable of forming whole OB subgroups.

\section{Summary and Conclusions}
OB associations with subgroups showing well defined age sequences 
and small internal age spreads
suggest a large-scale triggered formation scenario, presumably due to 
               supernova/wind driven shock waves.
Expanding, initially wind driven superbubbles around OB groups 
produce coherent large-scale ISM flows that form new clouds;
supernova shock waves then compress these newly formed clouds and 
trigger the formation of whole OB subgroups (several thousand stars) in
locations with suitable conditions.
Other triggering mechanisms (e.g. radiatively driven implosion)
may operate simultaneously, but
seem to form only small groups of stars and thus appear to be
secondary processes.

\end{document}